\documentclass[pra,aps,a4paper,showpacs,superscriptaddress,twocolumn]{revtex4}
\usepackage{amsfonts}
\usepackage{amsmath}
\usepackage{graphicx}
\usepackage{dcolumn}
\usepackage{bm}

\begin{document}
\title{Probing the quantum ground state of a spin-1 Bose-Einstein condensate with cavity transmission spectra}
\author{J.~M. Zhang}
\affiliation{Beijing National Laboratory for Condensed Matter
Physics, Institute of Physics, Chinese Academy of Sciences, Beijing
100080, China} \affiliation{Department of Physics, Henan Normal
University, Xinxiang 453007, China}
\author{S. Cui}
\affiliation{Department of Physics, Henan Normal University,
Xinxiang 453007, China}
\author{H. Jing}
\affiliation{Department of Physics, Henan Normal University,
Xinxiang 453007, China}
\author{D.~L. Zhou}
\affiliation{Beijing National Laboratory for Condensed Matter
Physics, Institute of Physics, Chinese Academy of Sciences, Beijing
100080, China}
\author{W.~M. Liu}
\affiliation{Beijing National Laboratory for Condensed Matter
Physics, Institute of Physics, Chinese Academy of Sciences, Beijing
100080, China}

\begin{abstract}
We propose to probe the quantum ground state of a spin-1
Bose-Einstein condensate with the transmission spectra of an optical
cavity. By choosing a circularly polarized cavity mode with an
appropriate frequency, we can realize coupling between the cavity
mode and the magnetization of the condensate. The cavity
transmission spectra then contain information of the magnetization
statistics of the condensate and thus can be used to distinguish the
ferromagnetic and antiferromagnetic quantum ground states. This
technique may also be useful for continuous observation of the spin
dynamics of a spinor Bose-Einstein condensate.
\end{abstract}

\maketitle

Spinor Bose-Einstein condensates (BECs) distinguish themselves from
their scalar counterparts by their much richer internal structures
and dynamics \cite{stamper_spinor,machida}. Based upon the single
mode approximation (SMA), which is valid in most cases, the quantum
eigenstates of a spinor BEC has been extensively and thoroughly
studied with sophisticated algebraic techniques
\cite{cklaw,Koashi,ho}. Especially, for the spin-1 case, which we
focus on in this paper, it is found that there are two possibilities
for the ground state. Depending on a single parameter, the ground
state can be either of an antiferromagnetic (AFM) or a ferromagnetic
(FM) nature. The two phases differ in their rotational
properties---the AFM ground state is a $SU(2)$ singlet and thus
isotropic \cite{even}; while the FM ground state is a $SU(2)$
multiplet heavily degenerate and thus can be well directed (up to
the uncertainty principle associated with the $SU(2)$ algebra). From
the point of view of quantum phase transition, these two phases
deserve to be distinguished experimentally for their own sake.

In this paper we propose a method based on cavity quantum
electrodynamics techniques to fulfill this object. The basic idea is
similar to that of Mekhov \textit{et al}. in Ref. \cite{ritsh_NP},
where they proposed to probe the superfluidity-Mott insulator
transition of cold atoms in optical lattices with cavity
transmission spectra. In the dispersive regime, atoms inside the
cavity shift the cavity frequency and in turn influence the cavity
transmission spectra. The superfluidity and Mott-insulator phases
differ in their atom number statistics and thus lead to drastically
different cavity transmission spectra, which can be used inversely
to differentiate the two phases. In their proposal, it is adequate
to treat the atoms in a two-level fashion with the atomic levels
structureless. However, here in our case, we have to go beyond the
two-level model and take into account the detailed and realistic
atomic level structure. In this way, we can retrieve the vector
polarizability \cite{bonin} of the atom that is dropped artificially
in the two-level model. It follows that by choosing a circularly
polarized cavity mode, the cavity photon couples not only to the
total atom number but also to the total magnetization. The latter
effect allows us to do our job, since the cavity transmission
spectra carry information of the magnetization statistics of the
BEC, which are absolutely different in the two phases.

We assume a spin-1 BEC of $N$ alkali atoms of a certain type is
loaded inside an optical cavity \cite{bec_cavity}. This system is
probed with a $\sigma_+$ polarized laser so that the relevant cavity
mode coupled with the atoms is also $\sigma_+$ polarized. The bare
frequency $\omega_c$ of this cavity mode is chosen to resolve the
alkali $D$ line (fine structure) and placed relatively near the
$D_1$ transition \cite{why_D1}. However, the detuning between the
$D_1$ transition and the cavity mode frequency is still large enough
not to resolve the hyperfine structures of the $D_1$ line. More
precisely, we have $\Delta_{HFS1}\ll |\omega_c-\omega_{D1}|\ll
|\omega_c-\omega_{D2}|$, where $\Delta_{HFS1}$ is the magnitude of
the hyperfine splitting of the $D_1$ line, $\omega_{D1}$ and
$\omega_{D2}$ are the average frequency of the hyperfine levels of
$D_1$ and $D_2$, respectively. In the appendix we show that the
effective Hamiltonian describing the atom-cavity photon interaction
is ($\hbar=1$ throughout in this paper)
\begin{equation}\label{atom photon interaction}
    H_{int}=\sum_{m}\left(1+\frac{m}{2}\right)U_0 \int dx u^2(x)
    \psi^\dagger_m(x)\psi_m(x)a^\dagger a.
\end{equation}
Here $\psi_m(x)$ ($m=0,\pm 1$) is the field annihilation operator
for an atom in the hyperfine state $|F=1,m\rangle$ at position $x$
(the quantization direction is along the cavity axis and is denoted
as the $z$-direction), and $a$ is the annihilation operator for the
cavity photon. The cavity mode function with its maximum normalized
to unity is denoted as $u(x)$. The parameter $U_0$ characterizes the
strength of the ac-Stark shifts of the atomic levels per photon, in
the viewpoint of the atoms; or alternatively, the magnitude of the
cavity frequency shift per atom, in the viewpoint of the photons.
Formally, we can write $U_0$ as $U_0=g_0^2/(\omega_c-\omega_{D1})$
by introducing an effective atom-photon coupling strength $g_0$,
which incorporates the overall effect of coupling with all the
hyperfine levels in the $D_1$ line.

Here some comments are in order. The main feature of Hamiltonian
(\ref{atom photon interaction}) is the spin dependence of the level
shifts, which can be interpreted in terms of a fictitious magnetic
field. In most experiments aiming to study intrinsic spin dynamics
of a spinor BEC, this effect is (and should be) avoided or minimized
by taking the laser beam to be linearly polarized or with a detuning
too large to resolve the $D$ line or both (an exception being Ref.
\cite{corwin}). However, for the purpose of continuous observation
of spin dynamics and spin magnetization, this effect proves to be
useful. Actually, it is exactly the same physics that underlies the
Faraday rotation spectroscopy developed recently for \textit{in
situ} observation of a spinor BEC \cite{higbie,liu,mueller}. For if
the cavity mode is $\sigma_-$ polarized, the plus sign in front of
$\frac{m}{2}$ in Eq.~(\ref{atom photon interaction}) should be
replaced with a minus sign. That is, the $\sigma_+$ and $\sigma_-$
polarized light respond differently to a given atomic sample---the
essence of the Faraday rotation effect.

For the spinor BEC, we assume the SMA is valid and rewrite the
atomic field operators as $\psi_m(x)=\phi(x)c_m$. Here $\phi(x)$
denotes the common spatial mode (normalized to unity) for the three
spin components and $c_m$ is the annihilation operator associated
with the corresponding spin component. Under the SMA, we can rewrite
Hamiltonian (\ref{atom photon interaction}) as
\begin{equation}\label{SMA h}
    H_{int}=U \left(\hat{N}+\frac{1}{2}\hat{M}\right) a^\dagger a,
\end{equation}
where $\hat{N}=\sum_m c_m^\dagger c_m$ is the total atom number and
$\hat{M}=\sum_m m c_m^\dagger c_m$ is the total magnetization. The
parameter $U\equiv U_0\int dx u^2(x)\phi^2(x)$ is proportional to
the overlap between the cavity mode function and the condensate mode
function. In general, the extension of a condensate is much larger
than the period of $u^2(x)$ in the axial direction while much
smaller than its waist in the transverse direction. Therefore, for a
condensate placed around the axis of the cavity, the integral can be
well approximated with $1/2$, i.e., $U\simeq U_0/2$. We note that
$\hat{N}$ and $\hat{M}$ commute with both $H_{int}$ and the
Hamiltonian of the spinor BEC itself. Therefore, we have a quantum
non-demolition measurement.

Assume the probe laser is of frequency $\omega_p$. The Heisenberg
equation (in the frame rotating at $\omega_p$) for the cavity mode
annihilation operator is
\begin{equation}\label{heisenberg}
    \dot{a}=-i\left(\omega_c-\omega_p+ U \left(\hat{N}+\frac{1}{2}\hat{M}\right)\right)a-\kappa a
    +\eta,
\end{equation}
where $\kappa$ is the damping rate of the cavity mode and $\eta$ the
driving amplitude. The damping term leads to the stationary
solutions of $a$ and the photon number \cite{ritsh_NP},
\begin{subequations}
\begin{eqnarray}
  a_{st}=\frac{\eta}{\kappa+i(\Delta+U\hat{M}/2)},  \quad\quad \\
  \hat{n}_{st}=a^\dagger_{st}a_{st}=\frac{\eta^2}{\kappa^2+(\Delta+U\hat{M}/2)^2},
\end{eqnarray}
\end{subequations}
where $\Delta=\omega_c+UN-\omega_p$ is the shifted cavity-probe
detuning. The expectation value of $\hat{n}_{st}$ over a prescribed
state of the spinor BEC is then
\begin{equation}\label{expectation value}
    n_{st}=\langle \hat{n}_{st} \rangle =\sum_{M=-N}^N f_M
    \frac{\eta^2}{\kappa^2+(\Delta+UM/2)^2},
\end{equation}
where $f_M$ is the probability for $\hat{M}$ taking value $M$ in the
prescribed state. We have $\sum_M f_M=1$, where the sum is over all
the possible values of $M$. As pointed out in Ref.~\cite{jmz}, the
total photon number contains incoherent contributions of a series of
coherent states, the $M$-th of which is of amplitude
$\eta/(\kappa+i(\Delta+UM/2))$ and of weight $f_M$. Mathematically,
we see from Eq.~(\ref{expectation value}) that $n_{st}$, as a
function of the detuning $\Delta$, is a weighted sum of a series of
Lorentz functions, which are evenly displaced with $U/2$ and of
full-width half-maximum $2\kappa$. The point is that the cavity
transmission spectra (proportional to $n_{st}$) contain information
of the atomic statistics embodied in $f_M$.

We now apply the formalism developed above to the two possible
phases of a spin-1 BEC. For a spin-1 BEC in the SMA, the Hamiltonian
accounting for the spin dynamics is $H_s=W\textbf{S}^2$
\cite{cklaw}, with the spin operator
$\textbf{S}\equiv(S_x,S_y,S_z)=c_m^\dagger
(S_x^{mn},S_y^{mn},S_z^{mn}) c_n$. The parameter $W$ is proportional
to the difference of the s-wave scattering lengthes in the two total
spin channels ($F=0$ or $2$). Obviously, the sign of $W$ determines
the nature of the ground state. If $W>0$, the ground state is of
total spin $S=0$ and given explicitly as \cite{Koashi,ho,even}
\begin{equation}\label{anti ground}
    |G\rangle_{AFM}=|S=0,M=0\rangle \propto
    \left(A^\dagger\right)^{N/2}|0\rangle,
\end{equation}
where the scalar (rotationally invariant) operator $A^\dagger\equiv
c_0^{\dagger 2}-2c_{+1}^\dagger c_{-1}^\dagger$. This state is a
singlet and also the unique ground state. For this AFM ground state,
the magnetization has no fluctuations at all. The only nonvanishing
term in Eq.~(\ref{expectation value}) is $M=0$. Therefore, by
scanning the probe frequency $\omega_p$, we get the intracavity
photon number as a single Lorentz function,
\begin{equation}\label{expectation value anti}
    n_{st} (\Delta)=
    \frac{\eta^2}{\kappa^2+\Delta^2},
\end{equation}
with a width of $2\kappa$ and maximum height of unity (in units of
$\eta^2/\kappa^2$, the maximum possible intracavity photon number).
Note that it is of the same shape as that of a bare cavity, though
the center is shifted by $UN$.

If $W<0$, the ground state is of total spin $S=N$ and $2N+1$ fold
degenerate. Explicitly, we have as a set of orthonormal basis
\cite{Koashi,ho},
\begin{equation}\label{ferro ground}
    |G\rangle_{FM}=|S=N,M\rangle\propto (S_-)^{N-M}(c_{+1}^\dagger)^N
    |0\rangle,
\end{equation}
where $S_-=\sqrt{2}(c_0^\dagger c_{+1}+c_{-1}^\dagger c_0)$ is the
lowering operator for $S_z$. These $2N+1$ states each have a
definite magnetization, thus the corresponding transmission spectra
is also a single Lorentz function (but the center of which may
vary). However, here we are most interested in the experimentally
most accessible states---angular momentum coherent states---states
that can be obtained by rotating the $z$-aligned state
$|N,N\rangle$,
\begin{equation}\label{angular momentum coherent state}
 e^{-i\theta\vec{\zeta}\cdot
    \vec{S}}|N,N\rangle=\frac{(x_{+1}c_{+1}^\dagger +x_0c_{0}^\dagger
    +x_{-1}c_{-1}^\dagger)^N}{\sqrt{N!}}|0\rangle.
\end{equation}
Here $\vec{\zeta}=(-\sin\phi,\cos\phi,0)$ perpendicular to the
$z$-axis is the axis of rotation, while $\theta$ is the angle of
rotation or inclination angle. The $x_i$'s are determined by
$e^{-i\theta\vec{\zeta}\cdot
    \vec{S}} c_{+1}^\dagger e^{i\theta\vec{\zeta}\cdot
    \vec{S}}=x_{+1}c_{+1}^\dagger +x_0c_{0}^\dagger
+x_{-1}c_{-1}^\dagger$. The state constructed in Eq.~(\ref{angular
momentum coherent state}) defines a direction $(\theta,\phi)$ in
that it is the eigenstate of $S_{n(\theta,\phi)}=\sin\theta\cos\phi
S_x + \sin\theta\sin\phi S_y +\cos\theta S_z$ with eigenvalue $N$.
These states can be readily prepared with a proper radio-frequency
(rf) pulse from the state $|N,N\rangle$ \cite{sengstock}. For such a
state, the weight coefficients of the Fock states
$|N_{+1},N_0,N_{-1}\rangle$ follow the triple-nominal distribution
\begin{equation}\label{triple distribution}
    P(N_{+1},N_0,N_{-1})=\frac{N!}{N_{+1}!N_0!N_{-1}!}p_{+1}^{N_{+1}}p_0^{N_{0}}p_{-1}^{N_{-1}},
\end{equation}
with $p_i=|x_i|^2$ and $\sum_i p_i=1$. Explicitly, we have
$(p_{+1},p_0,p_{-1})=(\cos^4\frac{\theta}{2},2\cos^2\frac{\theta}{2}\sin^2\frac{\theta}{2},\sin^4\frac{\theta}{2})$.
Note that $p_i$ are independent of the azimuth angle $\phi$. This is
reasonable since all angular momentum coherent states with the same
inclination angle $\theta$ are connected by rotations around the
$z$-axis. Therefore they share the same distribution
$P(N_{+1},N_0,N_{-1})$.

\begin{figure}[t]
\centering
\includegraphics[bb=15 15 330 325,width=0.45\textwidth]{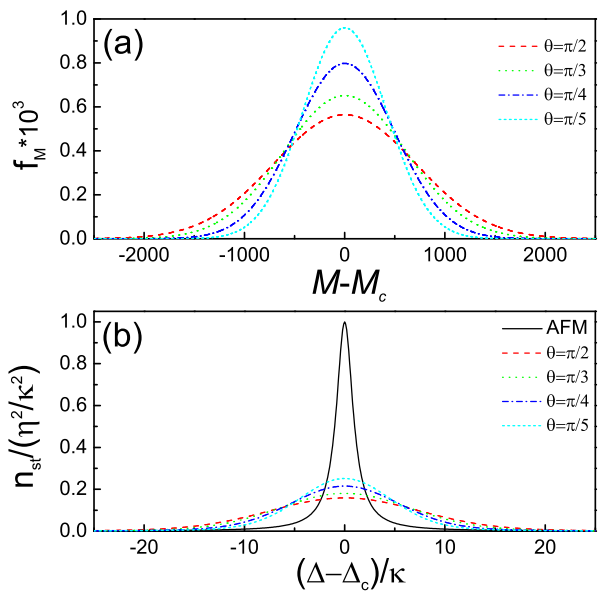}
\caption{\label{fig1} (Color online) (a) Probability distribution
function $f_M$ of the magnetization $M$ of ferromagnetic (FM) ground
states with different inclination angles. (b) Intracavity photon
number at steady state $n_{st}$ as a function of the detuning
$\Delta$. The solid line corresponds to the antiferromagnetic (AFM)
ground state. Other lines correspond to FM ground states in (a). The
parameters are $N=10^6$ and $U=0.01\kappa$. For the horizontal
labels, $M_c=N\cos\theta$ and $\Delta_c=-M_cU/2$ in the FM case,
while $M_c=0$ and $\Delta_c=0$ in the AFM case.}
\end{figure}

The probability distribution function $f_M$ can be generated in a
random walk process. Image a particle, initially at the origin,
performing random walks on the real axis. At each step, the particle
either moves rightward or leftward or just keeps still, with
probabilities $p_{+1}$, $p_{-1}$, and $p_0$, respectively. Then
$f_M$ is just the probability of arriving at $M$ after $N$ steps. In
other words, $M$ is the sum of $N$ independent random variables
$\{X_k,1\leq k \leq N \}$, the common distribution function of which
is $Pr(X_k=i)=p_i$, with $i=0$, $\pm 1$. This interpretation enables
us to calculate $f_M$ both analytically and numerically.
Analytically, for $N\gg 1$, by the central limit theorem, we can
approximate $f_M$ with the normal distribution
\begin{equation}\label{normal distribution}
    f_M=\frac{1}{\sqrt{2\pi}\sigma} \exp\left(-\frac{(M-M_c)^2}{2\sigma^2}
    \right),
\end{equation}
where $M_c=N\cos\theta$ and $\sigma=\sqrt{N}\sigma_0$ (here
$\sigma_0=\sqrt{\frac{1}{2}\sin^2\theta}$ being the variance of
$X_k$) are respectively the average value and variance of $M$.
Numerically, the characteristic function of $f_M$ is
$\chi(t)=(p_{+1}e^{+it}+p_0+p_{-1}e^{-it})^N$. By an inverse Fourier
transform, we get $f_M$ determined (it is just the coefficient of
$e^{iMt}$ in the expansion of $\chi(t)$). The two approaches yield
the same result up to a high precision. In Fig.~\ref{fig1}(a), we
show the distribution $f_M$ for four different inclination angles
with the total atom number $N=10^6$. We see the larger the absolute
value of $\sin\theta$, the larger the fluctuations of $M$. For all
cases, we see a clear $\sqrt{N}$ scaling of $\sigma$.

We calculate the intracavity photon number $n_{st}$ by approximating
$f_M$ with Eq.~(\ref{normal distribution}) and replacing the sum
with an integral. The integral can be further understood as the
convolution of the probability distribution function $f_M$ and a
Lorentz function. We can then use the convolution theorem to first
solve the Fourier transforms of the two functions respectively, form
their product, and then take an inverse Fourier transform. We get
\begin{eqnarray}
  n_{st}(\Delta) &=&\frac{\eta^2}{2\kappa}\int_{-\infty}^{+\infty} dt e^{it(\Delta-\Delta_c)}\times
  e^{-\kappa|t|-t^2U^2\sigma^2/8},\quad \label{trans final
  expression}
\end{eqnarray}
where $\Delta_c=-M_cU/2$. Obviously, the procedure can be done
inversely, i.e., from the cavity transmission spectra observed
experimentally, we can infer the distribution of the magnetization
by Fourier transforms.

We can read two facts from Eq.~(\ref{trans final expression}).
Firstly, the shape of $n_{st}$ as a function of $\Delta$ is shifted
by $\Delta_c$ from that depicted in Eq.~(\ref{expectation value
anti}). This overall shift is due to the average deviation of $M$
from zero and thus has a classical nature. Secondly, the shape of
$n_{st}$ is broadened compared to that in Eq.~(\ref{expectation
value anti}), which is due to the fluctuations of $M$ around the
mean and thus has a quantum nature. The shift and broadening are
appreciable if $(M_cU,\sigma U)\gg \kappa$, or in terms of $N$,
$(NU,\sqrt{N}U) \gg \kappa$. Currently, $g_0$ is about $2\pi \times
15$ MHz and the line width $\kappa$ of the cavity can be made as
small as $2\pi \times 1.0$ MHz
\cite{chapman,ritter,stamper-kurn_prl}. By taking the detuning
$\Delta_{ca}=\omega_c-\omega_{D1}$ to $2\pi \times 40$ GHz, we get
$U\sim 2\pi \times 2.8$ kHz. The atom number of a spinor BEC is
typically on the order of $10^6$. Thus we see the first condition
$NU\gg\kappa$ is well satisfied while the second condition
$\sqrt{N}U \gg \kappa$ is marginally satisfied. In the near future,
we expect $\kappa$ can be reduced by one order and then the second
condition can be well met also.

In Fig.~\ref{fig1}(b), we present the intracavity photon number at
steady state $n_{st}$ as a function of the detuning $\Delta$. For
the AFM ground state, the curve (solid line) is a simple Lorentzian
centered at $\Delta_c=0$ and with width $2\kappa$. This is the only
possibility. However, for the FM ground states, depending on the
inclination angle $\theta$, the center and width of the curve vary
widely (the figure may be a bit misleading, but note that we have
artificially shifted the curves to a common center; see the labels
of the horizontal axes). Especially, the broadening effect is
apparent compared to the AFM case (here we have
$\sqrt{N}U/\kappa=10$). Thus the AFM and FM ground states can be
distinguished on one hand by the different shapes of the
corresponding spectra \cite{shift} and on the other hand by their
responses to rf pulses (or a transverse magnetic field). Under the
action of rf pulses, the AFM ground state remains unaltered since it
is rotationally invariant, so the cavity transmission spectra are
unchanged. In contrast, a FM ground state can be tickled to other
directions and thus both the center and the width of the
transmission spectra will change.

An alternative and more convenient method may be just reversing the
polarization of the probe laser. Because the $\sigma_+$ and
$\sigma_-$ polarized cavity modes couple respectively to $\hat{M}$
and $-\hat{M}$, it is easy to see that for the AFM case, the
transmission spectra are unaffected; while for the FM case, the
transmission spectra are reflected to be centered at $-\Delta_c$.
The distance between the two centers is on the order of $NU$, which
is much larger than $\kappa$, therefore this effect can be readily
observed. This method makes evident the similarity between our
approach and the Faraday rotation spectroscopy.

In summary, we have shown that by choosing a circularly polarized
cavity mode, we can achieve coupling between the photon number with
the magnetization, i.e. the $z$-component of the total spin of the
spinor BEC. Full statistics of the magnetization of the spinor BEC
can be extracted from the cavity transmission spectra and this
allows us to discern the AFM and FM phases, which due to their
different rotational properties, possess absolutely different
magnetization statistics. Our discussion has been focused on the
thermodynamical properties of the BEC (ground state oriented).
However, it is clear that the technique can also be used for
observing the real-time evolution of a spinor BEC. Moreover, since
the condition $NU\gg \kappa$ can be readily fulfilled, the
sensitivity can be high.

It should be emphasized that our discussion has been greatly
simplified by neglecting the center-of-mass motion of the BEC and by
the SMA [without the two approximations, we cannot reduce
Eq.~(\ref{atom photon interaction}) to Eq.~(\ref{SMA h})]. These
approximations are reasonable if the probe is weak enough. For
future works, however, the strong probe case is worth investigation.
If the probe is so strong that the intracavity optical lattice
cannot be neglected, center-of-mass motion of the spinor BEC can be
excited. Moreover, since the optical lattice felt by different spin
components are different, a dynamical SMA is inappropriate. The
situation is further complicated by the back-action of the BEC on
the cavity mode. In the end, this system may exhibit many
interesting nonlinear effects \cite{jmz2}.

This work was supported by NSF of China under Grant Nos.~10874235,
60525417, and 10775176, NKBRSF of China under Grant
Nos.~2009CB930704, 2006CB921400, 2006CB921206, and 2006AA06Z104.
J.~M.~Z. would like to thank J. Ye for stimulating discussions.

\appendix*
\section{Effective Hamiltonian}
In this appendix we review the derivation of the effective
Hamiltonian (\ref{atom photon interaction}). The main concern is to
demonstrate the Zeeman-like effect. General theory about this effect
has actually been well established \cite{cho,jessen,cohen}. We
include the derivation here just to serve our concrete purpose. For
simplicity, we take the optical field to be classical,
generalization to the quantum case is straightforward. The alkali
atom in question is assumed to be in the $n^{2}S_{1/2}$ $F=1$
hyperfine ground state manifold. A laser with electric field
$\vec{E}(t)=\vec{E}\exp(-i\omega_f t)+c.c.$ couples the atom to the
$n^{2}P_{1/2}$ excited state manifold ($D_1$ line). Here it is
assumed that the frequency of the laser is chosen so that the fine
structure of the alkali $D$ line is well resolved while the
hyperfine structure is not. This is feasible since the fine
structure splitting is generally three or four orders larger in
magnitude than the hyperfine structure splitting for the alkali $D$
line. For instance, the fine splitting of the $D$ line of $^{87}$Rb
is $2\pi\times 7200$ GHz, while the hyperfine splitting of the $D_1$
line is $2\pi \times 814.5$ MHz.

We aim to derive an effective Hamiltonian confined in the $F=1$
ground state subspace. By second order perturbation theory and
taking the rotating wave approximation, we have \cite{common}
\begin{equation}\label{second order}
    H_{eff}=\frac{P_{g;F=1} (\vec{E}^*\cdot \vec{d})P_{D1}(\vec{d}\cdot \vec{E})P_{g;F=1}
    }{\omega_f-\omega_{D1}},
\end{equation}
where $\vec{d}$ is the electric dipole operator and $P_{g;F=1}$ and
$P_{D1}$ are the projection operators on the $F=1$ hyperfine ground
state manifold and $D1$ fine excited state manifold respectively.
That is, $P_{g;F=1}=\sum_m |n^{2}S_{1/2};F=1,m\rangle \langle
n^{2}S_{1/2};F=1,m|$ and $P_{D1}=\sum_{F,m} |n^{2}P_{1/2};F,m\rangle
\langle n^{2}P_{1/2};F,m|$, where the sum is over all the possible
values of $F$ and $m$. For the latter, we can also transfer from the
coupled representation to the uncoupled representation, i.e.,
equivalently we have
\begin{equation}\label{uncoupled}
    P_{D1}=P_{e;J=1/2}\otimes P_{I}.
\end{equation}
Here $P_{e;J=1/2}$ is purely an electronic projection operator onto
the $J=1/2$ excited state subspace while $P_I$ is purely a nuclear
projection operator. We can also define the project operator
\begin{equation}\label{ground fine}
    P_g=P_{g;J=1/2}\otimes P_I=P_{g;F=1}+P_{g;F=2},
\end{equation}
which projects onto the ground state manifold up to the fine
structure. Substituting (\ref{uncoupled}) and (\ref{ground fine})
into (\ref{second order}) and noting that $P_g P_{g;F=1}=P_{g;F=1}$,
we rewrite the effective Hamiltonian as
\begin{widetext}
\begin{eqnarray}   \label{effective h 2}
 H_{eff}&=&P_{g;F=1} \frac{P_{g} (\vec{E}^*\cdot \vec{d})P_{D1}(\vec{d}\cdot
    \vec{E})P_{g}}{\omega_f-\omega_{D1}} P_{g;F=1}  \nonumber\\
   &=& P_{g;F=1}\left( \frac{P_{g;J=1/2} (\vec{E}^*\cdot \vec{d})P_{e;J=1/2}(\vec{d}\cdot
    \vec{E})P_{g;J=1/2}}{\omega_f-\omega_{D1}}\otimes P_I\right)
    P_{g;F=1}.
\end{eqnarray}
\end{widetext}
Note that in the parentheses we have successfully separated the
electronic part from the nuclear part. This is important since the
external electric field couples with the electrons but not the
nucleus. We define $\vec{D}=P_{e;J=1/2} \vec{d} P_{g;J=1/2}$,
%\begin{equation}\label{D}
%
%\end{equation}
which is a vector operator (and so is $D^\dagger$). The electronic
part in the parentheses in (\ref{effective h 2}) is then in the form
$(\vec{E}^*\cdot \vec{D}^\dagger)(\vec{D}\cdot \vec{E})$, which can
be rewritten in the form $D_i^\dagger D_j E_i^* E_j$, the scalar
contraction of two rank-2 tensors, with the atomic and optical parts
separated. We can decompose the dyadic $D_i^\dagger D_j$ into
rank-$0$, $1$, $2$ irreducible tensors,
\begin{widetext}
\begin{equation}\label{irreducible}
    D_i^\dagger D_j=\frac{1}{3}\left(\vec{D}^\dagger\cdot
    \vec{D}\right)\delta_{ij}+\frac{1}{2}\left(D_i^\dagger D_j-D_j^\dagger D_i\right)+\left[
    \frac{1}{2}\left(D_i^\dagger D_j+D_j^\dagger D_i\right)-\frac{1}{3}\left(\vec{D}^\dagger\cdot
    \vec{D}\right)\delta_{ij}
    \right],
\end{equation}
\end{widetext}
with $E^*_i E_j$ done similarly. We note that since $D^\dagger D$
acts on a two-dimensional Hilbert space, the rank-2 irreducible part
vanishes identically. We can then contract the corresponding
irreducible tensors to form scalars. The result is
\begin{equation}\label{result}
    \frac{1}{3} \left(\vec{D}^\dagger\cdot
    \vec{D}\right)\left(\vec{E}^*\cdot
    \vec{E}\right)+\frac{1}{2}\left(\vec{D}^\dagger \times \vec{D}\right)\cdot
    \left(\vec{E}^*\times \vec{E}\right).
\end{equation}
Using the Wigner-Eckart theorem, we see that in the $J=1/2$ ground
state subspace the scalar operator $\vec{D}^\dagger\cdot \vec{D}$ is
proportional to the identity operator while the vector operator
$\vec{D}^\dagger \times \vec{D}$ to $\vec{J}$. Thus the above
equation is equivalent to
\begin{equation}\label{euivalent 1}
    |\vec{E}|^2\left(\alpha-i\beta (\vec{e}^*\times \vec{e})\cdot \vec{J}
    \right),
\end{equation}
with $\alpha$ and $\beta$ being some constants and
$\vec{e}=\vec{E}/|\vec{E}|$ being the laser polarization vector.
Substituting this result into (\ref{effective h 2}) and using again
the Wigner-Eckart theorem, we have finally
\begin{equation}\label{effective h final}
    H_{eff}=\frac{|\vec{E}|^2}{\omega_f-\omega_{D1}} \left(
    \alpha-i\gamma (\vec{e}^*\times \vec{e})\cdot \vec{F}
    \right),
\end{equation}
with $\gamma$ being some constant incorporating the Land\'e factor
\cite{grimm}. The first term describes a center-of-mass shift of the
ground state multiplet while the second term is in effect a Zeeman
term---the very effect we are seeking. Note that for a linearly
polarized laser the second term vanishes identically while for
$\sigma_+$ and $\sigma_-$ polarized lasers, the fictitious magnetic
fields point opposite.

Therefore we have shown how an effective Hamiltonian like (1) can
arise naturally under certain ciucumstances. To get the coefficients
$\alpha$ and $\gamma$ determined, it is better to resort to the
dipole matrix elements tabulated in Ref.~\cite{steck}.


\begin{thebibliography}{99}
\bibitem{even}
In this paper we assume the total atom number $N$ is even. For $N$
being odd, the AFM ground state is of total spin $S=1$ and thus not
strictly isotropic. However, that makes little difference or
difficulty to our proposal; see discussions below.
\bibitem{why_D1}
In contrast to many experiments (e.g. Refs.
\cite{chapman,ritter,stamper-kurn_prl}), here the $D_1$ line is
preferred to the $D_2$ line. The reason is that under the same
detuning, the $D_1$-scheme yields the same coupling with the
magnetization as the $D_2$-scheme, while the coupling with the
density is only one half, so the disturbance introduced is smaller.
\bibitem{shift}
Experimentally the total atom number $N$ cannot be determined
precisely, so we cannot distinguish the AFM and FM ground states
just by the shift of the spectra.
\bibitem{common}
The condition that we do not resolve the hyperfine structure of the
$D_1$ line allows us to take a common denominator.
\end{thebibliography}
\end{document}